# 3D forest semantic segmentation using multispectral LiDAR and 3D deep learning

Narges Takhtkeshha[1,2]*, Lauris Bocaux[1], Lassi Ruoppa[3], Fabio Remondino[1], Gottfried Mandlburger[2], Antero Kukko[3], Juha Hyyppä[3]

[1] 3D Optical Metrology (3DOM) unit, Bruno Kessler Foundation (FBK), Trento, Italy;
Email: ntakhtkeshha@fbk.eu, lbocaux@fbk.eu, remondino@fbk.eu
[2] Department of Geodesy and Geoinformation, TU Wien, Vienna, Austria; Email: gottfried.mandlburger@geo.tuwien.ac.at
[3] Department of Photogrammetry and Remote Sensing, Finnish Geospatial Research Institute, National Land Survey of Finland, Espoo, Finland; Email: lassi.ruoppa@nls.fi, antero.kukko@nls.fi, juha.hyyppa@nls.fi

**Abstract**
Conservation and decision-making regarding forest resources necessitate regular forest inventory. Light detection and ranging (LiDAR) in laser scanning systems has gained significant attention over the past two decades as a remote and non-destructive solution to streamline the labor-intensive and time-consuming procedure of forest inventory. Advanced multispectral (MS) LiDAR systems simultaneously acquire three-dimensional (3D) spatial and spectral information across multiple wavelengths of the electromagnetic spectrum. Consequently, MS-LiDAR technology enables the estimation of both the biochemical and biophysical characteristics of forests. Forest component segmentation is crucial for forest inventory. The synergistic use of spatial and spectral laser information has proven to be beneficial for achieving precise forest semantic segmentation. On the other hand, 3D deep learning (DL) models have demonstrated outstanding performance in point cloud segmentation compared to traditional methods. Thus, this study aims to investigate the potential of MS-LiDAR data, captured by the HeliALS system, providing high-density multispectral point clouds to segment forests into six components: ground, low vegetation, trunks, branches, foliage, and woody debris. Three point-wise 3D deep learning models and one machine learning model, including kernel point convolution (KPConv), superpoint transformer (SPT), point transformer V3 (PTv3), and random forest (RF), are implemented. Our experiments confirm the superior accuracy of the KPConv model. Additionally, various geometric and spectral feature vector scenarios are examined. The highest accuracy is achieved by feeding all three wavelengths (1550 nm, 905 nm, and 532 nm) as the initial features into the deep learning model, resulting in improvements of 33.73% and 32.35% in mean intersection over union (mIoU) and in mean accuracy (mAcc), respectively. This study highlights the excellent potential of multispectral LiDAR for improving the accuracy in fully automated forest component segmentation.

1. **Introduction**

By providing a holistic understanding of the forest ecosystem, forest inventory supports sustainable management practices that balance economic, ecological, and social values. However, forest measurements are still primarily collected through labor-intensive in-situ methods using simple tools such as calipers, measuring tapes, and hypsometers. These direct measurement techniques are time-consuming and prone to human error. Over the past two decades, very high-density terrestrial laser scanners (TLS) have garnered significant attention for their ability to collect detailed 3D forest structural data with millimeter-level accuracy, enabling rapid, automated, and periodic forest inventories (Tanaka et al., 1998; Liang et al., 2016). TLS systems cover significantly smaller areas compared to airborne laser scanning (ALS), and there is a need for multiple scans to map even small test plots. TLS typically speeds up plot measurements, but a large amount of work is needed in the processing of the data. Conversely, recent unmanned laser scanners (ULS) can capture dense point clouds of entire tree structures in a relatively short time, presenting substantial potential to improve the accuracy and efficiency of detailed tree measurements and forest inventory (Jaakkola et al. 2010; Wallace et al., 2012). Accurate measurement of certain tree attributes often depends on one or two specific parts of the tree. Tree foliage plays a critical role in controlling forest productivity, health, and responses to climate change, as it directly influences photosynthesis and evapotranspiration. Foliage profiles are also of particular interest for biodiversity conservation. Trunks and branches are essential for obtaining quantitative structural models (QSMs) of trees and estimating the timber strength (Winberg et al., 2023; Pehkonen et al., 2025). Tree trunks contribute to the estimation of tree's diameter at breast height (DBH), biomass, stem volume, and stem quality (Levick et al., 2016), while identifying branches is necessary to exclude them from leaf area index (LAI) and biomass estimation processes (Calders et al., 2015). The presence of woody debris serves as a biodiversity indicator, as many threatened species rely on decaying wood for habitat. Additionally, forest component segmentation facilitates individual tree segmentation and tree modelling (Heinzel et al., 2018). Therefore, separating forest component is a useful pre-processing step for all forestry applications development whether they predict or directly measure tree attributes from point clouds. Additionally, such information would be beneficial even for simultaneous localization and mapping (SLAM) processes in mobile laser scanners to avoid mismatching errors caused by under-storey and foliage points (Kaijaluoto et al., 2022).

Currently, forest semantic segmentation primarily relies on the geometric information derived from LiDAR data. Some studies have explored the impact of single-wavelength information from traditional monochromatic (mono) LiDAR systems on forest component segmentation and demonstrated its advantage (Windrim et al., 2020; Kaijaluoto et al., 2022). Since the back-scattered and captured light by the sensor varies depending on the material hit by the laser, incorporating laser intensity data can enhance segmentation accuracy (Morel et al., 2020). The advent of MS-LiDAR technology, which operates across multiple channels, enables the simultaneous collection of spatial and multispectral information (Takhtkeshha et al., 2024a). As forest components exhibit distinct optical properties at different operating wavelengths, MS-LiDAR data can help with distinguishing between various forest elements (Disney, 2018). The effects of mono- and multispectral information on forest component segmentation is discussed in detail in Subsection 2.1. While the geometric information from LiDAR data facilitates the extraction of structural metrics, the spectral information provided by MS-LiDAR systems not only enhances the accuracy of these characteristics but also offers new critical insights into tree health monitoring. By identifying parameters such as water content and biochemical composition, MS-LiDAR systems enable more reliable and comprehensive forest inventory practices.

To date, numerous innovative approaches have been proposed for semantically segmenting forest environments. Tao et al. (2015) horizontally sliced point clouds to detect circle-like shapes for separating wood and leaves. Ferrara et al. (2018) applied density-based spatial clustering (DBSCAN) to classify wood and non-wood points while Heinzel et al. (2018) employed mathematical morphology techniques to identify tree trunks. However, these model-based methods are designed for a limited range of tree species and architectures, making them less robust when applied to diverse tree structures. Machine learning (ML) algorithms using hand-crafted features such as linearity, planarity, scattering and surface curvature have been explored in several studies to estimate DBH, stem curvature, and stem volume (Yun et al., 2016; Ma et al., 2016; Hyyppä et al., 2020). Wang et al. (2017) compared four popular ML classifiers—support vector machine (SVM), naive bayes (NB), random forest, and gaussian mixture model (GMM)—for distinguishing wood from leaves using TLS data. Their findings identified RF as the most effective model while emphasizing the critical role of density-related features. Despite these advancements, dense understory vegetation and complex forest structures continue to challenge the generalizability of most conventional approaches. Even for human experts, accurately distinguishing trunk boundaries, foliage initiation points, and other classes can be difficult. This underscores the need for more advanced segmentation algorithms capable of handling diverse forest types, particularly evergreen tropical forests. DL models have shown exceptional promise in image segmentation tasks. To adapt these models to point cloud data and address challenges associated with processing irregular structures, specialized point-wise deep learning architectures have been developed since 2017 (Ruizhongtai Qi et al., 2017). Unlike ML, the notable advantage of DL models is their ability to automatically extract and select features. Recent studies have begun to employ DL models for segmenting various forest components (Morel et al., 2020; Windrim et al., 2020; Krisanski et al., 2021a; Krisanski et al., 2021b; Kaijaluoto et al., 2022; Shao et al., 2023; Xiang et al., 2024; Liang et al., 2024). These approaches unveiled the significant potential of DL models. A detailed review of DL-based studies is provided in Subsection 2.2. Although deep learning models have been applied in some studies for MS-LiDAR semantic segmentation (Takhtkeshha et al., 2024b; Reichler et al., 2024), to the best of our knowledge, they have not yet been used for forest MS-LiDAR.

To take full advantage of MS-LiDAR technology for precise forest inventory, firstly, forest semantic segmentation is a mandatory task. Thus, the primary objective of this paper is to introduce a reliable and automated forest semantic segmentation pipeline capable of handling structurally diverse and complex forest point clouds. Building on existing research and leveraging recent advancements in data acquisition and processing, we present a deep learning-based forest semantic segmentation pipeline utilizing high-density, three-wavelength multispectral LiDAR data. This work focuses on segmenting six forest components: ground, low vegetation, trunks, branches, foliage, and woody debris. The innovative aspects of the work are:

- The investigation of the impact of multi-wavelength (more than two) LiDAR data on forest component segmentation;
- The use of deep learning methods for the semantic segmentation of forest multispectral point clouds;
- The segmentation of forest point clouds into six forest components;
- The analysis of the influence of vegetation indices on the initial feature vector for DL-based forest semantic segmentation;
- The release of the first publicly available forest MS-LiDAR benchmark dataset (once the paper be accepted).

## 2. Related Work

### 2.1 Laser Radiometric Information in Forestry

To date, MS-LiDAR technology has demonstrated significant potential across diverse ecological applications, including enhancing the detection and segmentation of individual trees (Huo and Lindberg, 2020; Dai et al., 2018), identifying invasive tree species (Mielczarek et al., 2022), classifying tree species (Budei et al., 2017; Hakula et al., 2023), and estimating tree physiological and biochemical properties such as moisture content and foliar nitrogen (Gaulton et al., 2010; Gaulton et al., 2013; Eitel et al., 2014; Wallace et al., 2014; Junttila et al., 2018; Goodbody et al., 2020; Maltamo et al., 2020).

Table 1. Summary of deep learning-based forest component segmentation studies using laser scanning in the literature; the best method is bold.

| References | LiDAR data | Forest classes | Methodology(s) | Spectral effect | Remarks |
|---|---|---|---|---|---|
| Lu et al., 2025 | ULS | 5 (Ground, low vegetation, stem, live branches, and woody branches) | $^{3D}$ Sen-net | NA | Open data and code |
| Liang et al., 2024 | TLS | 3 (Trunk, branches, and foliage) | $^{3D}$ **Point transformer**, DGCNN, PointNet++ | NA | Open data (ForestSemantic); private code |
| Xiang et al., 2024 | ULS | 5 (Ground, low vegetation, stem, live branches, and dead branches) | $^{3D}$ ForAINet | -0.3% | The negative spectral effect could be due to combining two mono-spectral scanners from different regions. Open data and code |
| Shao et al., 2023 | ULS | 5 (Ground, debris, trunks, crowns, and others) | $^{3D}$ Superpoint Graph | NA | Private data and code |
| Kaijaluoto et al., 2022 | MLS | 4 (Ground, understory, trunk, and foliage) | $^{3D}$ **RandLA-Net** and $^{2D}$ LSSegNet | +1.3% | Open data, private code |
| Krisanski et al., 2021a | TLS | 5 (Terrain, vegetation, coarse woody debris, and stem) | $^{3D}$ PointNet++ | NA | Open code, private data |
| Krisanski et al., 2021b | TLS, MLS, ALS, and ULS | 5 (Terrain, vegetation, coarse woody debris, and stem) | $^{3D}$ Modified PointNet++ | NA | Open data and code |
| Windrim et al., 2020 | ALS | 2 (Stem and foliage) | $^{3D}$ Fully convolutional network | +2.9% | Private data and code |
|  |  |  | $^{3D}$ PointNet | +2.3% |  |
| Morel et al., 2020 | TLS | 2 (Leaf and wood) | $^{3D}$ Improved PointNet++ | NA | Private data and code |

Depending on the wavelength of the scanner, the intensity of the reflected LiDAR pulse can facilitate the differentiation of return signals resulting from interactions with various materials (Tao et al., 2015). Li et al. (2013) illustrated the potential of integrating two monochromatic laser scanners operating at 1548 nm and 1064 nm to separate leaves from trunks and branches. They observed that at SWIR wavelengths, leaves exhibit approximately half the brightness of trunks due to liquid water absorption. Conversely, at the NIR range of the electromagnetic spectrum, leaves and trunks displayed roughly equal brightness. The advantages of multi-wavelength systems for distinguishing wood from leaves and improving tree parameter estimation have been further validated by Danson et al. (2014), Howe et al. (2015), and Li et al. (2018). These studies utilized dual-wavelength (SWIR and NIR) full-waveform TLS systems, such as the Salford Advanced Laser Canopy Analyser (SALCA) and the dual-wavelength Echidna LiDAR (DWEL), alongside traditional methods like thresholding and random forest classification. The synergetic use of spatial and spectral information not only improves classification accuracy but also reduces variance across multiple vegetation types (Li et al., 2018). The application of spectral information in forestry is not limited to just multi-wavelength LiDAR systems.

Even incorporating a single wavelength spectral information into forest semantic segmentation tasks has demonstrated accuracy enhancement by up to 34% (Zhu et al., 2017; Windrim et al., 2020; Kaijaluoto et al., 2022). Béland et al. (2014) demonstrated that applying a threshold in the 1535–1550 nm spectral range effectively distinguished between leaf and woody materials. This advantage arises because solid targets, such as wood, produce sharply peaked return pulses compared to foliage, enabling radiometric information to aid in differentiating them. Although the impact of mono- and dual-wavelength LiDAR intensities on forest component segmentation has been investigated, to the best of our knowledge, no studies have explored forest semantic segmentation using more than two wavelengths. Besides, former studies are TLS-based. Notably, prior research employing MS-LiDAR systems has focused on segmenting forest laser data into woody and non-woody elements. In contrast, our study considers six forest classes, enhancing the 3D forest inventory.

## 2.2 Deep Learning for Forest Semantic Segmentation

As outlined in Section 2.1, the potential of DL models has remained unexplored in previous studies on forest semantic segmentation using MS-LiDAR data. Since this paper proposes a novel DL-based approach for forestry multispectral point cloud segmentation, this section reviews prior research on forest component segmentation utilizing DL, as summarized in Table 1. According to Table 1, most DL models employed in this domain are based on PointNet++ (Qi et al., 2017), while other notable models include Superpoint Graphs (Landrieu and Simonovsky, 2018), RandLA-Net (Hu et al., 2019), point transformer (Zhao et al., 2021), DGCNN (Wang et al., 2018), and fully convolutional networks. The additional attributes of multispectral point clouds make graph-based DL models computationally expensive. The superiority of point transformer over PointNet++ and DGCNN has been demonstrated by Liang et al. (2024). Kaijaluoto et al. (2022) proposed a novel method for real-time semantic segmentation of mobile laser scanning (MLS) forest point clouds by utilizing raw, non-georeferenced 2D laser scanner profiles formatted as a sequence of 2D rasters. Each raster series included range, reflectance, and echo deviation data at a single scan line level. Their U-Net-based 2D convolutional neural network (LSSegNet) demonstrated slightly lower accuracy compared to the point-wise RandLA-Net model. More recently, Lu et al. (2025) presented a KPConv-based model called Sen-net for semantic segmentation of ULS data across five forest components, demonstrating superior performance compared to state-of-the-art deep learning models such as RandLA-Net.

Among the reviewed studies, only three out of nine DL-based papers have examined the effect of laser intensity. Importantly, none of them have explored the impact of multispectral laser information. Windrim et al. (2020) and Kaijaluoto et al. (2022) substantiated that incorporating single-wavelength LiDAR intensity data can enhance stem-foliage separation accuracy by up to 2.9%. Conversely, Xiang et al. (2024) reported that while involving monochromatic LiDAR intensity improved segmentation accuracy for ground and low vegetation, it reduced the accuracy for stem and dead branch segmentation. A slight decrease is also observed for live branch segmentation accuracy. The -0.3% decrease in mIoU of segmentation may be attributed to the characteristics of the FOR-Instance dataset, which comprises a mix of mono-spectral data from five countries that are collected using two different sensors, acting at different wavelengths—RIEGL VUX-1 UAV (1550 nm) and RIEGL miniVUX-1 UAV (905 nm).

While most existing studies have focused on binary leaf-wood separation, some recent research, such as that by Shao et al. (2023), Xiang et al. (2024), and Lu et al. (2025), has expanded the classification to include multiple forest components. Shao et al. (2023) identified ground, debris, trunks, crowns, and other elements, whereas Xiang et al. (2024) and Lu et al. (2025) categorized forest structures into ground, low vegetation, stems, live branches, and dead branches. In this paper, we selected three state-of-the-art deep learning models and one machine learning algorithm for segmenting six forest components: KPConv (Thomas et al., 2019), SPT (Robert et al., 2023), PTv3 (Wu et al., 2024), and RF. According to the literature, RF has been proved as the best machine learning classifier for forest component segmentation (Wang et al., 2017). Superpoint transformer and point transformer V3 are new 3D deep learning algorithms that have not yet been used for forest semantic segmentation. KPConv has shown promising results in forest component segmentation in the recent study by Lu et al. (2025). It is important to note that PointNet++ has not been implemented in this study because PointNet++ is among the first 3D deep learning models, and the superiority of the point transformer over PointNet++ for forest component segmentation has already been demonstrated by Liang et al. (2024). The details of the semantic segmentation process are thoroughly presented in Subsections 4.1- 4.6. Detailed information on these DL models is presented in Section 4.

## 3. Study Area and Dataset

The case studies are four biodiverse natural forest regions located in Espoonlahti in southern Finland (Figure 1a). In Espoonlahti area, there are close to 20 different tree species available, the most common are pine, spruce, birch, maple, aspen, rowan, oak, lime, and alder. High-density point cloud dataset is acquired in July 2023 in leaf-on condition by the helicopter-mounted HeliALS multispectral LiDAR system developed by the Finnish Geospatial Research Institute (FGI). This MS-LiDAR system is a combination of three single-wavelength laser scanners, namely VUX-1HA, miniVUX-1DL, and VQ-840-G (Hakula et al. 2023). Thus, this system can capture 3D data in three wavelengths: 1550 nm, 905 nm, and 532 nm. Detailed information on the HeliALS system is reported in Table 2.

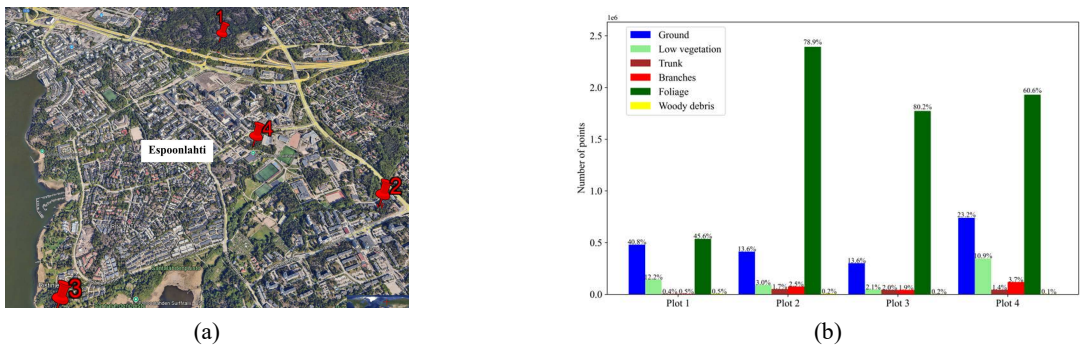

Figure 1. Study plots (a) in Espoonlahti, Finland, and the corresponding distribution of forest components (b).

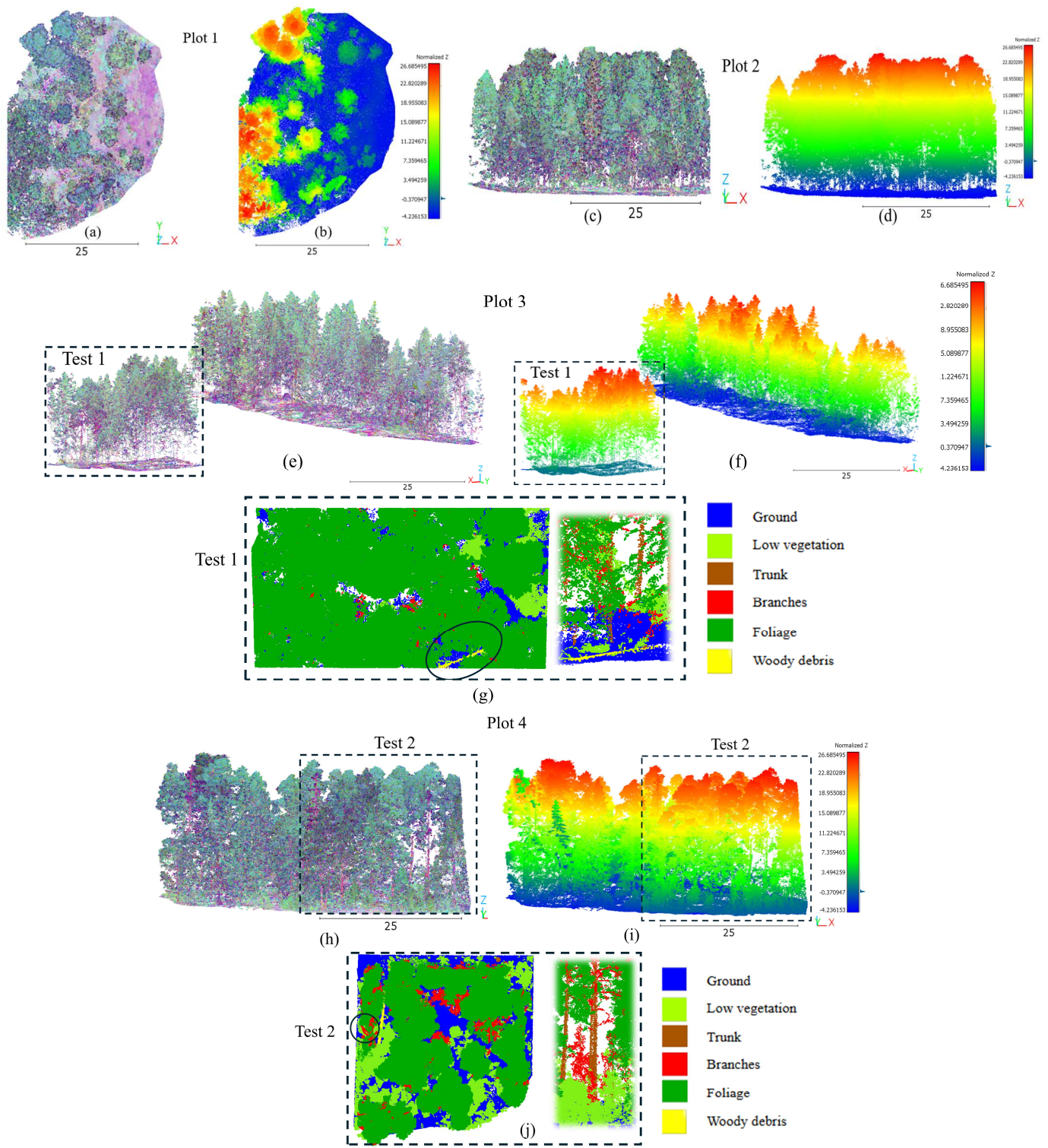

Figure 2. Multispectral forest point clouds (a, c, e, h), visualized using pseudo-coloring based on the intensity values of the SWIR, NIR, and green scanners mapped to the red, green, and blue channels, respectively. The corresponding normalized height are shown in (b, d, f, i). Subfigures (g) and (j) present the manually generated ground truth for the test areas.

Ground truth data are created manually in six classes (ground, low vegetation, trunk, branches, foliage, and woody debris) using CloudCompare v2.13.0's segmentation tool to train and evaluate DL models. A total of 9601494 points are annotated. The dataset is divided into training and test sets with a ratio of 4:1. Because of the limited availability of training data and the complexity of annotating forest point clouds across six classes, validation data is not included in this study. Due to the complex nature of forest ecosystems, the classes within our dataset are not uniformly represented (Figure 1b). On average, woody debris, trunks, and branches account for only 0.23%, 1.51%, and 2.52% of the dataset, respectively. In comparison, low vegetation covers 6.54% of the forest, while 20.14% of the points correspond to the forest floor (ground). Most of the points (69.08%) are attributed to foliage. Hence, we have a highly class-imbalanced dataset. The multispectral point cloud dataset used in this study is presented in Figure 2 and is publicly available as the first forest MS-LiDAR benchmark dataset for use in future research.

Table 2. Specification of the multispectral LiDAR dataset over the forestry area.

| Scanner | VUX-1HA | miniVUX-1DL | VQ-840-G |
|---|---|---|---|
| Wavelength (nm) | 1550 | 905 | 532 |
| Average point density (point/m$^2$) | 530 | 163 | 604 |
| Pulse repetition rate (kHz) | 1017 | 100 | 200 |
| Beam divergence (mrad) | 0.5 | 0.5 × 1.6 | 1 |
| Flight altitude (m) | | ∼ 100 | |

### 3.1 Multispectral LiDAR vs. Monochromatic LiDAR

Figure 3 depicts the geometry difference between the multispectral and monochromatic point clouds. While monochromatic laser scanner components of some MS-LiDAR systems, like Optech Titan and RIEGL VQ-1560i-DW, have roughly the same point density, the varying pulse repetition rates of individual monochromatic laser scanners in the HeliALS system result in differences in the geometries of individual laser wavelengths. Among the monochromatic laser scanners, VUX-1HA provides the most accurate representation of individual trees, though it lacks full corresponding points for the low vegetation and woody debris. Conversely, the VQ-840-G offers superior geometric representation of the low vegetation and woody debris, but has sparse points for branches, ground, and particularly trunks. The miniVUX-1DL performs the worst in scanning forest components. According to Figure 3d, the multispectral HeliALS system not only enhances the spectral richness of point clouds but also provides the most comprehensive representation of all forest classes thanks to its three scanners. To more thoroughly assess the advantages of MS-LiDAR systems over conventional mono-LiDAR scanners, we conducted an ablation study using different single-reflectance values. The results are presented in Subsection 5.2.

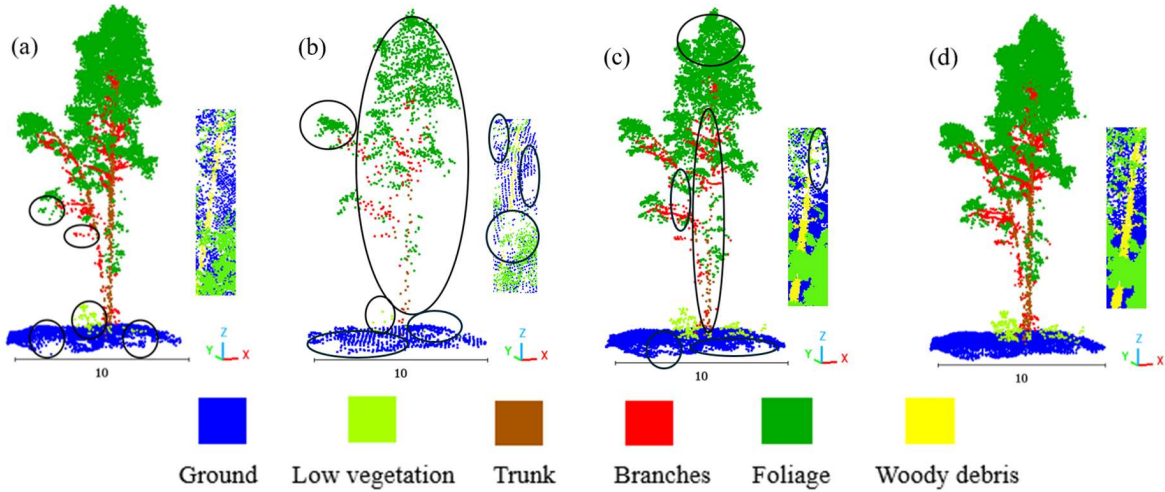

Figure 3. Surveyed geometry of forest components: mono-scanners (VUX-1HA (a), miniVUX-1DL (b), VQ840-G (c)) vs. MS-LiDAR system (d).

## 4. Methodology

Traditional machine learning methods usually rely on extracting representative geometric features such as verticality, linearity, planarity, sphericity, and principal components to facilitate the forest point cloud semantic segmentation. As previously mentioned, this study aims to explore the potential of radiometric information of MS-LiDAR data for boosting forest semantic segmentation. The spectral histograms of the six forest classes are illustrated in Figure 4. According to these subfigures, all six forest components possess unique multispectral distributions. Notably, although the spectral behaviours of low vegetation and foliage are roughly similar in the Green and NIR wavelengths, they act differently in SWIR wavelength. As a result, we propose a spatial-multispectral pipeline for forest semantic segmentation using 3D deep learning and MS-LiDAR data which synergistically integrates geometric and multispectral laser information. We employ laser reference values instead of amplitude values, as the former are range-corrected (Pfennigbauer and Ullrich, 2010).

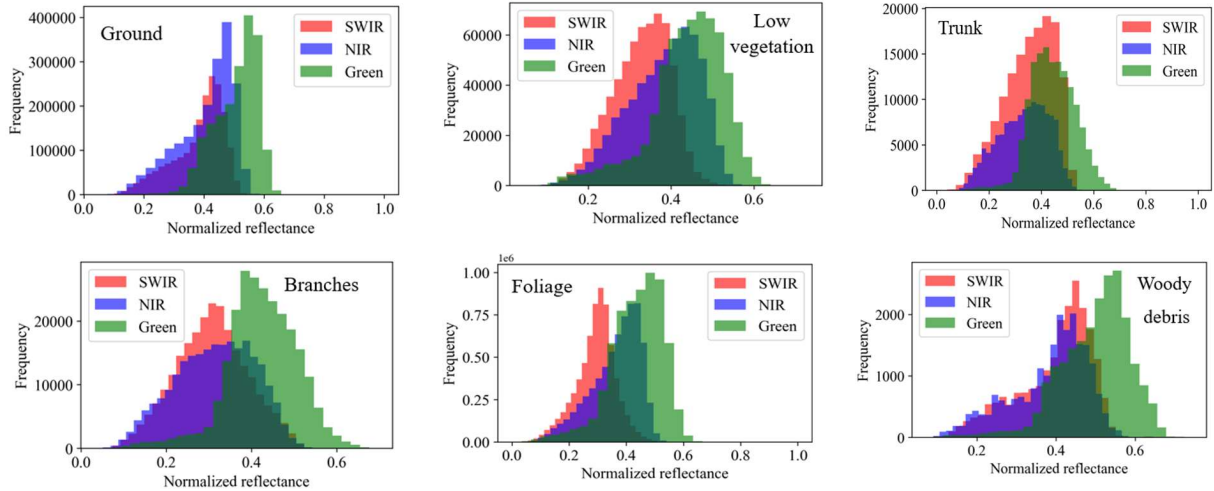

Figure 4. Distribution of reflectance values of forest components across different laser wavelengths. Reflectance values are scaled between 0 and 1 for visualization purposes.

Figure 5 provides an overview of the proposed methodology. In the first step, outliers in the point clouds from each monochromatic laser scanner are removed using the SOR (statistical outlier filter) tool in CloudCompare, with default parameters. Subsequently, multispectral point clouds are generated by merging the individual monochromatic laser data using the preMergeChannelsPointclouds module in OPALS software (Mandlburger et al., 2009; Pfeifer et al., 2014). The number of neighbors is configured to one, with a search radius of 0.25 meters in 3D space. Points for which complete spectral information (i.e., all three wavelengths) cannot be obtained from neighboring points are discarded. The average point density of the resulting multispectral point clouds is 1198.19 points/m², approximately twice as high as the densest mono-scanner (VQ-840-G). Subsequently, the coordinates and reflectance values of the multispectral point clouds are fed into deep learning models. We implemented three state-of-the-art 3D deep learning models—KPConv, SPT, and PTv3 — along with established random forest machine learning algorithm to compare them for spatial-multispectral forest semantic segmentation.

Moreover, we investigated the effect of MS LiDAR information by considering various initial feature vector scenarios, including solely reliance on 3D coordinates, single-channel reflectance, dual-channel reflectance, reflectance values of all channels, and the inclusion of a vegetation index (VI). Detailed information on the integration of vegetation indices is provided in Section 4.1.

The results are validated using established metrics, including overall accuracy (OA), mean accuracy, intersection over union (IoU), and mean IoU. These metrics are based on the true positives (TP), true negatives (TN), false positives (FP), and false negatives (FN). TP represents the number of points that belong to a specific forest component in the real world and are correctly predicted by the model. TN refers to the number of points that the model correctly identifies as not belonging to that forest component. FP indicates the number of points that the model incorrectly classifies as belonging to a certain forest component. FN are the points that the model fails to identify correctly as part of a certain forest class. Furthermore, to compare deep learning models and initial feature vectors tailored for forest inventory purposes, we propose a new evaluation metric that assigns twice the importance to key forest components—trunks, branches, foliage, and woody debris. Given that mIoU is the primary metric for assessing semantic segmentation, we introduce weighted IoU (wIoU). In this approach, the weight of each class is doubled for trunks, branches, foliage, and woody debris to better reflect their significance in forest inventory. All experiments are conducted on a Linux Mint 21.3 machine, equipped with a 32-core AMD EPYC 9354 CPU and an NVIDIA L40 GPU with 48.31 GB of onboard memory. The entire code for our study is developed in Python and is publicly available at https://github.com/3DOM-FBK/3D-forest-semantic-segmentation.

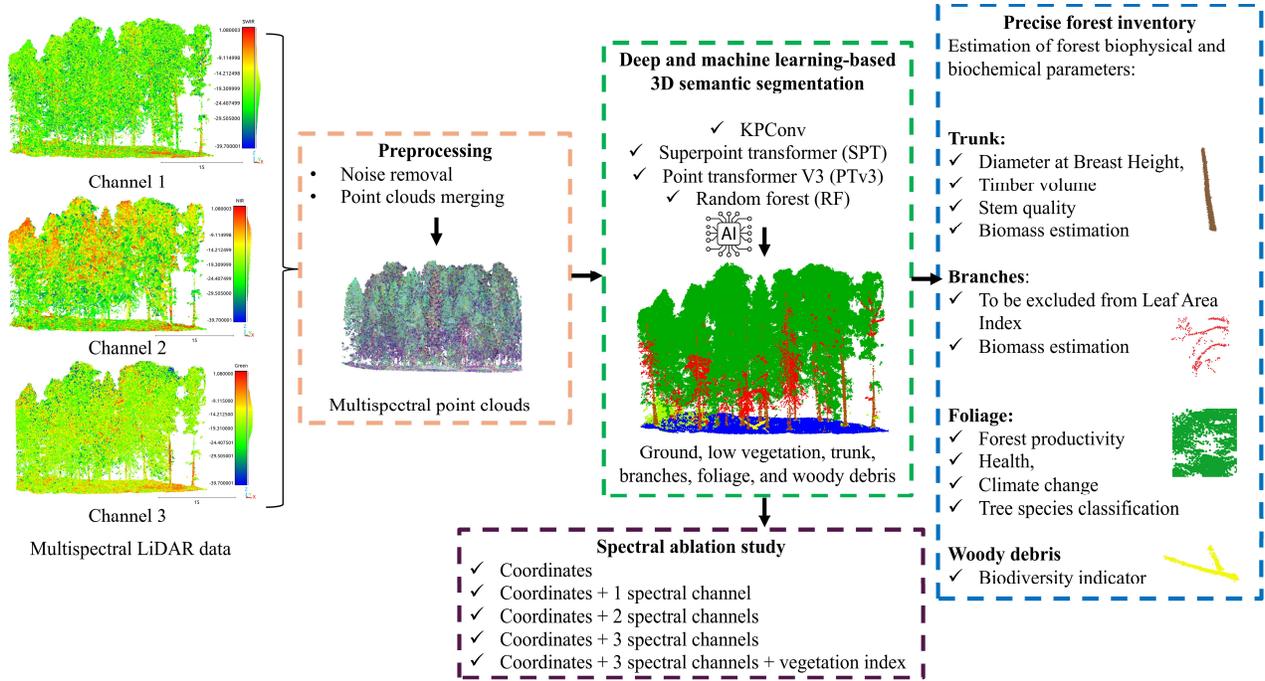

Figure 5. Proposed pipeline for precise forest inventory based on semantic segmentation of MS-LiDAR data.

### 4.1 Common Considerations in Semantic Segmentation

To normalize the features, the point clouds are centered along the planimetric axes (X and Y) according to Equations 1 and 2. Heights are normalized by subtracting each point's height from the local minimum height using OPALS software and then making sure that elevations are positive by subtracting the minimum value from height values (Equation 3). Other spectral and geometric features are normalized by applying a combination of robust (outlier-resistant) scaling and min-max scaling as described in Equation 4. In this formula, f denotes the spectral feature, and IQR (interquartile difference) is the difference between the 75th and 25th quantiles.

$$X = X - mean(X) \qquad (1)$$

$$Y = Y - mean(Y) \qquad (2)$$

$$Z = Z - min(Z) \qquad (3)$$

$$f\_normalized = \frac{f - mdeian(f)}{IQR(f)} \qquad (4)$$

where $IQR(f) = Q_{75}(f) - Q_{25}(f)$

$$f\_normalized = \frac{f\_normalized - min(f\_normalized)}{max(f\_normalized) - min(f\_normalized)}$$

Most advanced 3D deep learning models have built-in data augmentation tools. In some cases, both geometric and spectral augmentations are considered. For example, KPConv randomly drops a specific percentage of spectral information and introduces noise. However, since reflectance values correspond to the characteristics of the target being hit and remain stable under varying illumination conditions thanks to active LiDAR sensors—unlike images—typical radiometric augmentation techniques such as jittering, value dropping, and random noise can negatively impact reflectance values. As a result, these techniques may cause models to rely more on geometric features. Therefore, we deactivated such radiometric augmentations in the models we used.

### 4.2 KPConv

KPConv is a distinctive architecture that positions convolutional weights in space using kernel points, rather than relying on multi-layer perceptron (MLP) encodings (Thomas et al., 2019). We utilized the open-source PyTorch implementation of KPConv[1]. To mitigate class imbalance, class frequency-based weighting is applied to the Cross-Entropy loss function during segmentation. For parameterizing the KPConv model in the forest semantic segmentation task, we empirically set the initial subsampling grid size to 2 cm, the input sphere radius to 4 cm, the number of kernel points to 30, and the batch size to 6. Furthermore, we configured the convolution behavior to Gaussian and the aggregation function to "closest". The model is trained for 300 epochs with a learning rate of 0.01. The default stochastic gradient descent (SGD) optimizer is used. The remaining parameters are set to their default values.

### 4.3 Superpoint Transformer

The PyTorch-based Superpoint Transformer[2] is publicly available. This model incorporates a fast implementation of the cut pursuit algorithm for partitioning point clouds into a hierarchical superpoint structure, along with a self-attention mechanism that captures relationships between superpoints across multiple scales (Robert et al., 2023). This model stands out for its high computational efficiency, which is particularly important when processing high-density and high-dimensional multispectral point clouds. To tackle the class imbalance issue, we weighed the Cross-Entropy loss by computing a weighted average of the per-sample losses, encouraging the model to pay more attention to underrepresented classes. Additionally, we weighed the sampling of our data when creating batches to oversample under-represented classes and subsample over-represented ones.

SPT utilizes a set of point feature vectors for hierarchical partitioning of the point cloud and training. By default, these vectors include several hand-crafted features that are automatically extracted by SPT. In our approach, we selected normalized height, reflectance at three wavelengths, verticality, linearity, planarity, scattering, and curvature as the point features used for both partitioning the point cloud into superpoints and for training. The voxel resolution is set to 1 cm. The number of nearest neighbors is set to 25, with a search radius of 1 meter. The most critical and sensitive parameters in SPT are regularization, spatial_weight, and cutoff, which serve as inputs to the Cut Pursuit partitioning algorithm. We empirically set the values of regularization, spatial_weight, and cutoff to [0.3,0.6,0.8], [1e-1, 1e-1, 1e-1], and [10, 10, 10], respectively, based on observations from example datasets. The model is trained for 400 epochs with a learning rate of 0.1, a weight decay of 0.01, and the default AdamW optimizer.

### 4.4 Point Transformer V3

Point transformer V3 represents the latest advancement in the point transformer model series. A key innovation of this architecture is its use of the self-attention mechanism for point cloud segmentation, first introduced by Zhao et al. (2021). Compared to its predecessors, PTv3 is specifically designed to address the traditional trade-off between accuracy and efficiency by removing the reliance on relative positional encoding and replacing complex attention patch interaction mechanisms with a streamlined approach optimized for serialized point clouds (Wu et al., 2024). We used open-source PTv3[3]. To address the class imbalance problem, we employed the Lovász-softmax loss function. The training is conducted over 800 epochs using the default AdamW optimizer with a batch size of 4, a learning rate of 0.01, and a weight decay of 0.05.

### 4.5 Random Forest

Random forest classifier has demonstrated superior performance compared to other machine learning models in the literature. Considering the features chosen by Wang et al., 2017 and our experimental results, we selected 15 hand-crafted features extracted using CloudCompare's "compute geometric features" tool, experimentally applied at 1 m local neighbourhood radius. These features include linearity, planarity, verticality, eigenentropy, surface variation, anisotropy, omnivariance, sum of eigenvalues, second eigenvalue, first and second elements of principal component analysis (PCA). In addition, reflectance values of the three wavelengths and $NDVI_{NIR-SWIR}$ as supplementary features. Detailed information on the reasons for considering $NDVI_{NIR-SWIR}$ as the vegetation index is provided in Subsection 4.6. The main parameters for the RF classifier are set as follows: number of estimators = 200, maximum depth = 50, maximum features = log2, minimum samples split = 2, and minimum samples leaf = 10. The class weight is set to "balanced" to deal with the imbalanced dataset.

### 4.6 Integration of Vegetation Index

As the RIEGL reflectance values are in dB, to calculate vegetation indices, we first converted reflectance values to linear units by Equation 10 (Pfennigbauer and Ullrich, 2010). We choose an appropriate vegetation index from computed five vegetation indices frequently used in the literature and compared their sensitivity to forest components. Using reflectance values, we calculated the normalized difference vegetation indices ($NDVI_{NIR-Green}$ and $NDVI_{NIR-SWIR}$), chlorophyll vegetation index (CVI), green ratio vegetation index (GRVI), and green difference vegetation index (GDVI), as expressed in Equations 5-10 (Luo et al., 2022; Goodbody et al., 2020; Pan et al., 2019). Based on the results presented in Figure 6, the $NDVI_{NIR-SWIR}$ demonstrates superior capability in distinguishing forest components. Hence, $NDVI_{NIR-SWIR}$ is considered the selected vegetation index to be fed into the best-performing deep learning model, alongside the reflectance values.

---

[1] https://github.com/HuguesTHOMAS/KPConv-PyTorch
[2] https://github.com/drprojects/superpoint_transformer
[3] https://github.com/Pointcept/PointTransformerV3

$$\text{reflectance}_{\text{linear}} = 10^{\frac{\text{reflectance}_{db}}{10}} \quad (5)$$

$$\text{NDVI}_{\text{NIR-Green}} = \frac{(\text{NIR} - \text{Green})}{(\text{NIR} + \text{Green})} \quad (6)$$

$$\text{NDVI}_{\text{NIR-SWIR}} = \frac{(\text{NIR} - \text{SWIR})}{(\text{NIR} + \text{SWIR})} \quad (7)$$

$$\text{CVI} = \frac{\text{NIR} \times \text{Green}}{\text{SWIR}^2} \quad (8)$$

$$\text{GRVI} = \frac{\text{NIR}}{\text{Green}} \quad (9)$$

$$\text{GDVI} = \text{NIR} - \text{Green} \quad (10)$$

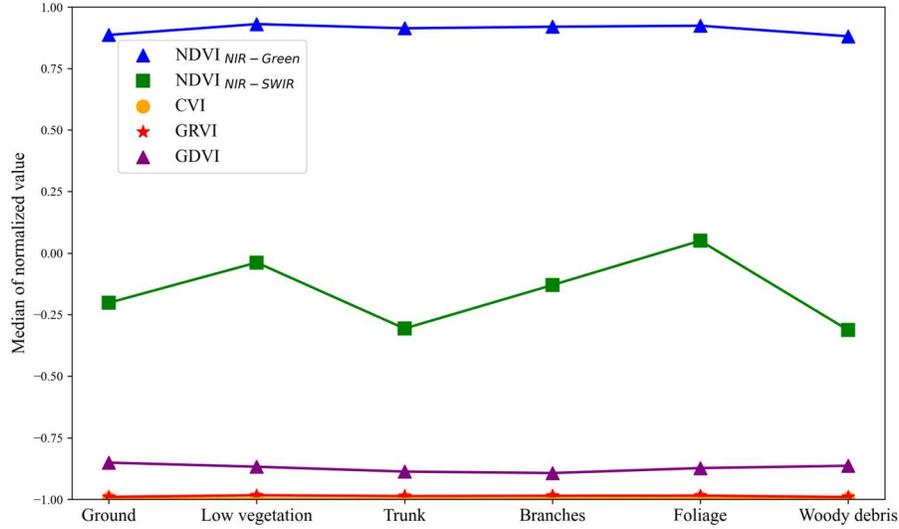

Figure 6. The sensitivity of the different vegetation indices to distinguishing forest components.

## 5. Results

### 5.1 Performance of Different Deep Learning and Machine Learning Models

Figure 7 reports the accuracy of forest semantic segmentation achieved by the mentioned DL and ML classifiers. According to this figure, KPConv demonstrates a substantial performance advantage over the other classifiers. Deep learning models outperform the traditional random forest machine learning model, improving wIoU by approximately 49.24%. Among the deep learning models, KPConv achieved the highest accuracy, followed by PTv3 and SPT, which exhibited the next highest accuracies, respectively. The pretrained KPConv weights are available in our GitHub repository and can be used as initial weights for training in future studies. Since KPConv outperforms the other classifiers, we consider this model for further spectral ablation study.

Figure 8 presents the confusion matrices for multispectral point cloud segmentation using KPConv across two test plots, and Figure 9 illustrates the segmentation results produced by KPConv. In both test plots, segmentation accuracy exceeds 70% across all classes, except for branches in the first test plot. In Figure 9, each row depicts a segmented forest element, with colors indicating ground truth labels. According to the achieved results, foliage, ground, low vegetation, and woody debris are segmented with high (more than 92%) TP. Moreover, trunk points are adequately segmented as trunks (TP > 77%), so that trunks are correctly detected in general. Most false positives in ground segmentation are due to misclassification as low vegetation. Typically, low points are assumed to be ground points, and the above points can be low vegetation. Therefore, this error reflects the difficulty in distinguishing the forest floor when it is obscured by low vegetation. In the first test plot, low vegetation is predominantly misclassified as ground, while in the second plot, it is often confused with foliage. This is likely due to the similar spectral characteristics of foliage and low vegetation, particularly when foliage appears at lower canopy levels. Trunks are frequently misclassified as foliage or branches, as these elements usually emerge directly from the trunk. Branches are commonly confused with foliage, and vice versa, due to their irregular spatial distribution among one another. Woody debris is mostly misclassified as ground, followed by low vegetation, as all occur at comparable height levels.

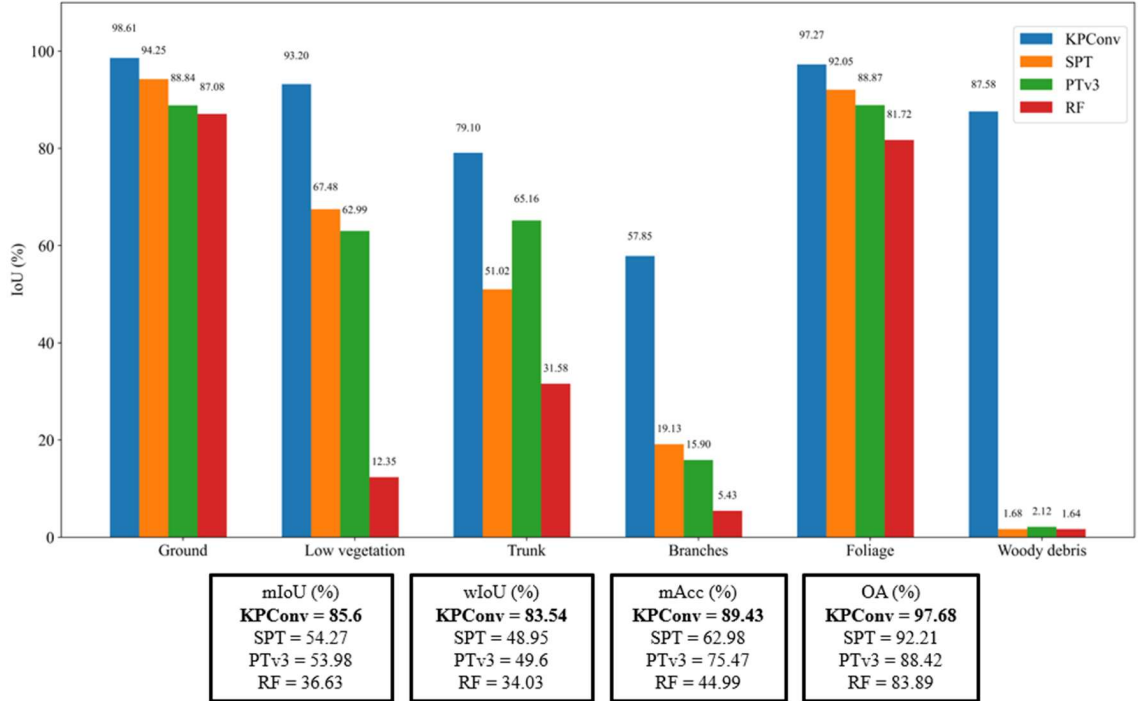

Figure 7. Multispectral forest point clouds semantic segmentation using different classifiers.

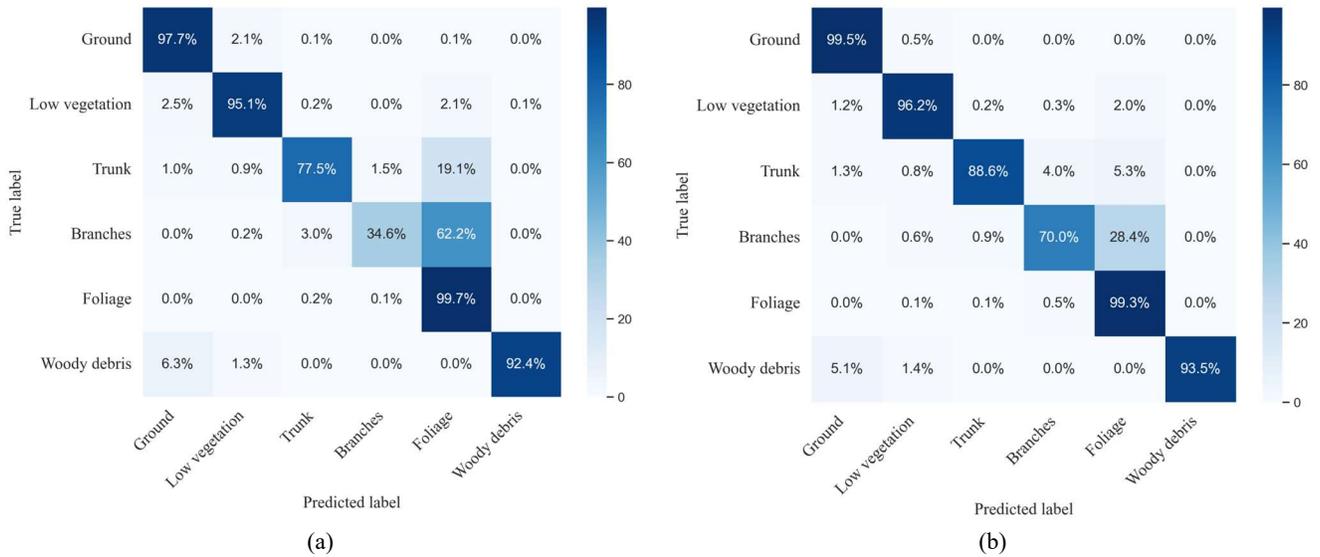

(a)                                                    (b)

Figure 8. Confusion matrices of the KPConv-based (the best classifier according to Figure 7) multispectral forest point clouds semantic segmentation; (a) test plot 1, (b) test plot 2.

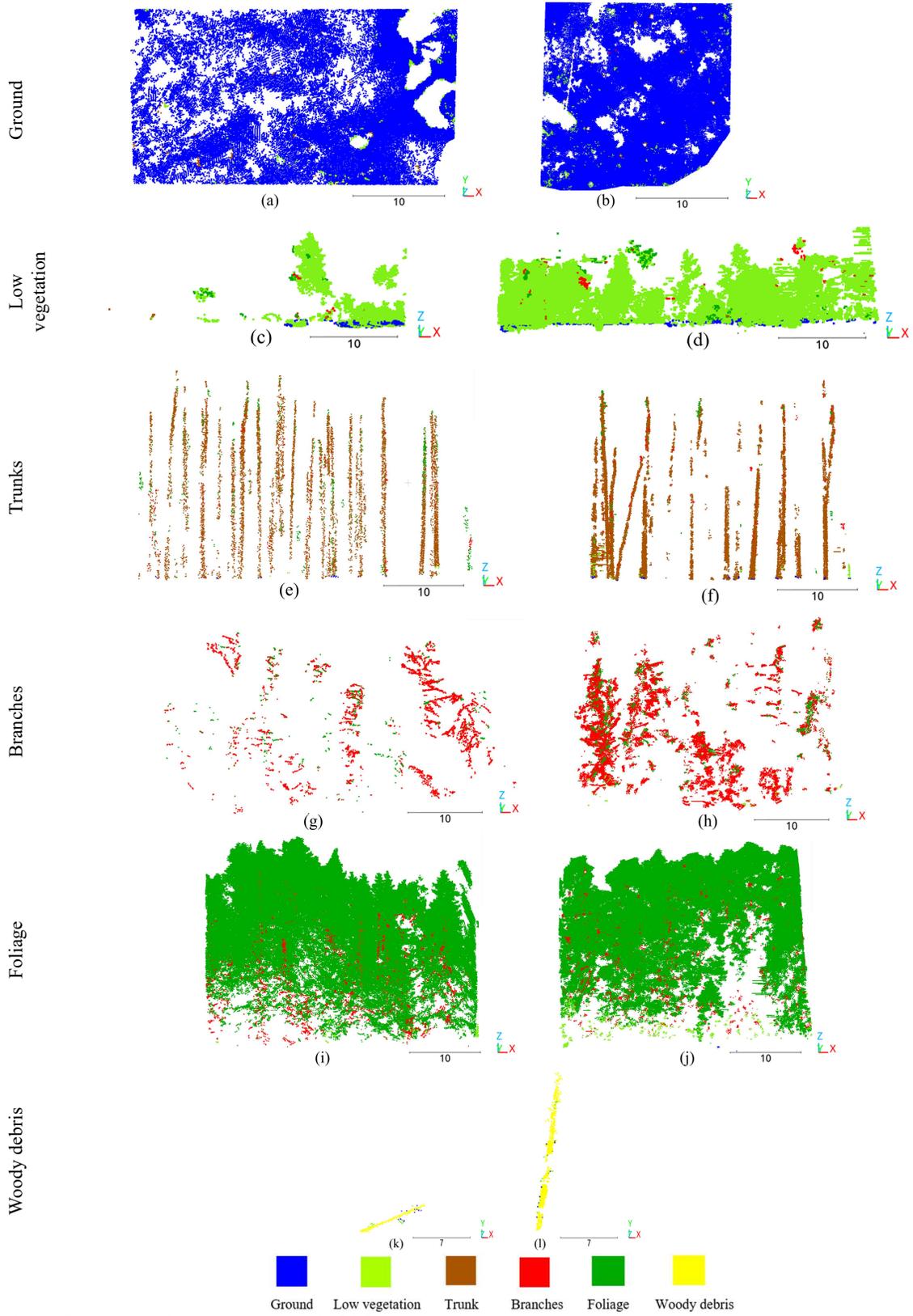

Figure 9. Visual results of multispectral LiDAR-based forest component segmentation using the best classifier. (a, c, e, g, i, k) correspond to test plot 1, while (b, d, f, h, j, l) are associated with test plot 2. Point clouds of detected components are colored based on the ground truth.

## 5.2 Multispectral Data Effect

We investigated the effect of incorporating spectral information on forest semantic segmentation through an ablation study, beginning with using all three available reflectance values in conjunction with NDVI $_{NIR-SWIR}$ as the vegetation index. The best-performing model, KPConv (see Figure 7), is employed for this spectral ablation study, and the corresponding results are presented in Table 3. Our experiments demonstrate that the incorporation of spectral information consistently improves the mAcc. While the segmentation performance for low vegetation and foliage improved across all scenarios, the inclusion of the NIR wavelength in the NIR and NIR + green combinations resulted in a marginal decline in ground segmentation accuracy, with a decrease of up to 0.31%. Trunk segmentation accuracy substantially improved with the addition of spectral information, except when SWIR and NIR are used together, which resulted in a slight decline of 0.54%. In contrast, branch segmentation accuracy decreased by up to 1.66 % when using NIR, SWIR + NIR, and SWIR + green combinations. Coordinate-based information alone yielded satisfactory results for detecting woody debris. Nevertheless, the synergistic use of SWIR and NIR, and especially the combination of all three wavelengths (NIR, SWIR, and Green), significantly enhanced woody debris segmentation accuracy by as much as 49.43%. Among all spectral configurations, the combination of all three wavelengths emerged as the most effective, yielding the highest overall improvement across all classes, with gains of 33.09% in wIoU and 32.35% in mAcc. Conversely, the combination of SWIR and green wavelengths resulted in the poorest spectral performance. Although it enhanced the mAcc by 4.06%, it led to a 2.35% decrease in the wIoU.

Table 3. Spectral ablation study on forest semantic segmentation based on the best-performing model (KPConv).

| Feature vector | IoU (%) | | | | | | mIoU (%) | wIoU (%) | mAcc (%) | OA (%) |
|---|---|---|---|---|---|---|---|---|---|---|
| | Ground | Low vegetation | Trunk | Branches | Foliage | Woody debris | | | | |
| Coordinates | 92.54 | 25.41 | 60.87 | 5.64 | 88.59 | 38.15 | 51.87 | 50.45 | 57.08 | 89.82 |
| +SWIR | 92.70 | 28.46 | 64.87 | 11.12 | 88.92 | 26.60 | 52.11 | 50.42 | 63.20 | 89.67 |
| +NIR | 92.50 | 35.55 | 64.31 | 3.98 | 89.29 | 29.59 | 52.54 | 50.24 | 62.10 | 90.30 |
| +Green | 92.60 | 36.47 | 63.10 | 6.35 | 89.44 | 28.30 | 52.71 | 50.35 | 62.81 | 90.59 |
| +SWIR + NIR | 92.87 | 33.77 | 60.33 | 3.77 | 89.13 | 38.53 | 53.07 | 51.02 | 58.74 | 90.34 |
| +SWIR + Green | 92.81 | 27.90 | 63.19 | 4.47 | 88.77 | 23.73 | 50.15 | 48.10 | 61.14 | 89.95 |
| +NIR + Green | 92.23 | 40.19 | 65.08 | 7.55 | 89.80 | 22.75 | 52.93 | 50.28 | 64.08 | 90.66 |
| +SWIR + NIR + Green | **98.61** | **93.20** | **79.09** | **57.85** | **97.27** | **87.58** | **85.60** | **83.54** | **89.43** | **97.68** |
| +SWIR +NIR +Green + VI | 92.58 | 29.11 | 63.35 | 5.77 | 88.79 | 33.44 | 52.17 | 50.44 | 61.68 | 89.98 |

## 6. Discussion

The efficiency of different DL and ML models for forest semantic sgemnation across six classes compared to each other in this study. Among the four classifiers implemented, KPConv surpassed the others with a 33.94% improvement in wIoU. Notably, KPConv demonstrated strong performance in segmenting complex forest components, including branches and woody debris. This aligns with findings by Robert et al. (2023), who reported that KPConv outperformed SPT in capturing intricate local geometries and improved segmentation of classes such as poles, fences, and vegetation in urban environments—further supporting our results. Nevertheless, KPConv demonstrated the poorest time efficiency, taking approximately ten times longer than SPT and PTv3. PTv3 outperformed SPT in segmenting trunk and woody debris classes. However, for the remaining classes, SPT achieved the second highest score after KPConv model. It is worth noting that we made considerable efforts to fairly allocate time for hyperparameter tuning across all models. However, there remains room for further improvement, particularly in identifying more optimal hyperparameters for the SPT, PTv3, and RF models.

According to Table 3, incorporating additional spectral information does not necessarily lead to improved performance. Notably, the combination of all three wavelengths with the vegetation index resulted in a lower wIoU compared to both the SWIR + NIR combination and the scenario of using all three wavelengths. On the other hand, the mono-spectral SWIR and green wavelengths showed greater accuracy than the bi-spectral combination of SWIR and green. Therefore, selecting the optimal spectral configuration is tricky and depends on the specific target classes of interest. To visually examine the effect of spectral configuration on forest semantic segmentation, we compare multispectral (including three wavelengths), bi-spectral, mono-spectral, and non-spectral point clouds in two zoomed plots. The combination of SWIR and NIR and the use of SWIR alone are selected as the bi-spectral and mono-spectral configurations, respectively, as they demonstrated the highest accuracy according to Table 3. As shown in Figure 10, the transition to multispectral point clouds resulted in a noticeable reduction in segmentation errors. Remarkably, the segmentation accuracy for ground, low vegetation, and foliage consistently improved. In both zoomed-in plots, while trunks are accurately segmented across all configurations, the mono-spectral setup outperforms the bi-spectral and coordinate-only configurations in segmenting branches and woody debris.

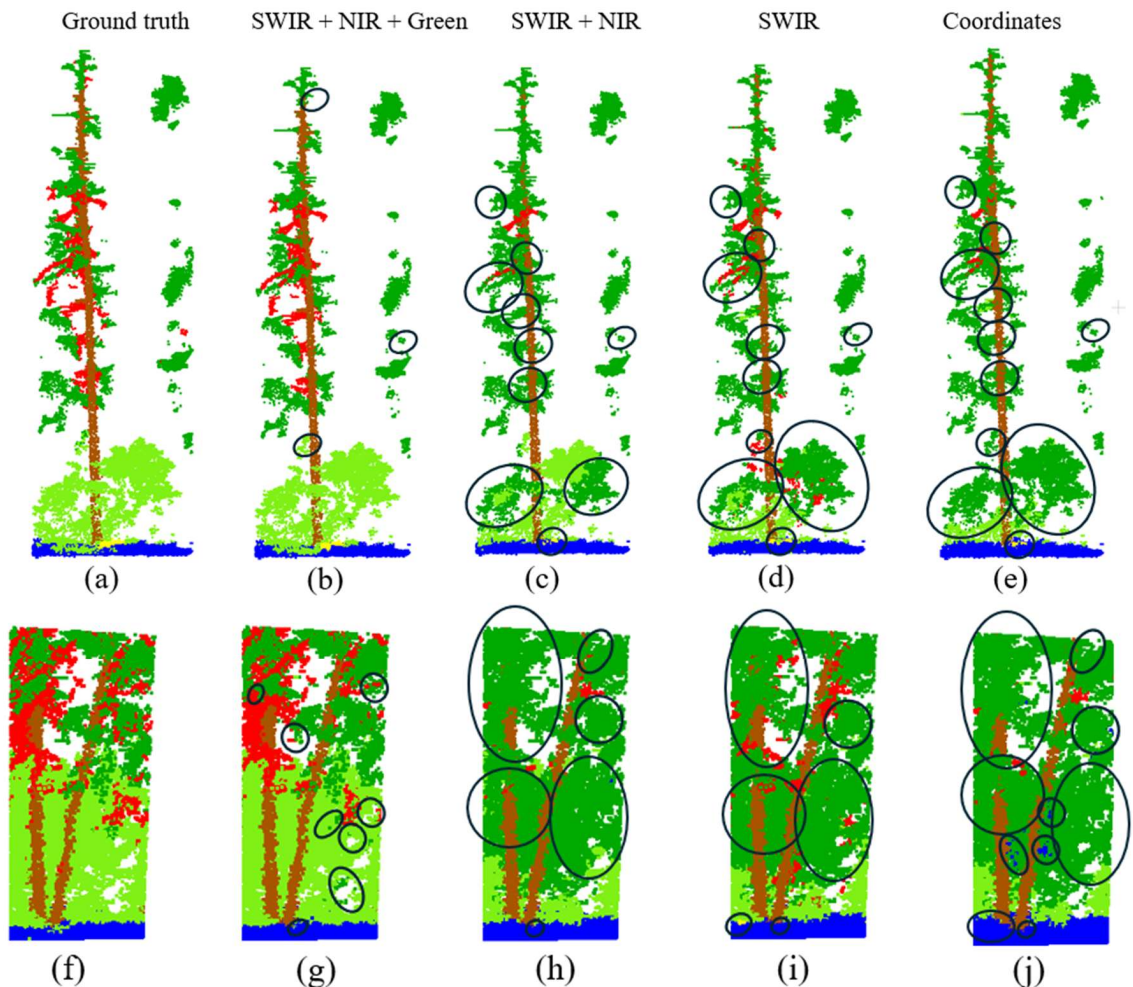

Figure 10. Visual comparison of multispectral (b and g), bi-spectral (SWIR and NIR; c and h), mono-spectral (SWIR; d and i), and coordinates only (e and j) point cloud data for forest component segmentation.

## 7. Conclusions

Although tree species reflect light differently, our experiments demonstrate the significant potential of laser reflectance in three wavelengths to enhance the semantic segmentation of forest point clouds by 33.09% in wIoU compared to solely using the coordinates of point clouds. Notably, our results show that even using the single-spectral information of a mono-LiDAR system boosts forest semantic segmentation, encouraging the deployment of laser reflectance, which has mostly received limited attention in current LiDAR forestry studies. Since MS-LiDAR simultaneously captures both 3D spatial and spectral information, and semantic segmentation serves as a foundational step in automated forest inventory, our proposed methodology presents a promising avenue for precise, single-source, and automated forest management. By improving the segmentation accuracy of forest components, MS-LiDAR data not only support more accurate and advanced measurements of biophysical attributes such as DBH, stem curve, and branching structure, but also enable the estimation of biochemical properties, including leaf water content, forest health, productivity status, and species classification.

Our proposed pipeline yields promising results for forest component segmentation, which can be further enhanced through post-processing techniques. For example, the segmented trunks can be utilized as initial points for defining stem shape using well-established geometrical methods. Besides, by detecting the trunks, the proposed framework streamlines the important and challenging task of individual tree segmentation. Integrating laser scanners with precise and AI-powered forest segmentation pipelines, like the one proposed, can significantly accelerate 3D forest inventory by reducing point cloud volume through the removal of unwanted forest components as specified by the end user.

In this study, we considered six forest components: ground, low vegetation, trunks, branches, foliage, and woody debris. Particular attention should be given to first-order branches in future research, as their location influences trunk and wood quality. In addition, a limited number of point clouds are annotated and used for training and evaluation of the results. The generalizability of the trained model on unseen datasets, as well as potential improvements using additional training data should be explored in future research. Moreover, our published pre-trained deep learning model weights, presented for both mono- and multispectral LiDAR, can facilitate future studies by serving as initial weights.

We utilized range-corrected reflectance values provided by RIEGL's V-Line laser scanners. Although these reflectance values yield promising results in forest semantic segmentation, it is imperative to explore robust, incidence angle–independent radiometric calibration solutions, particularly for complex multi-target environments such as forest environments in future studies. Incorporating radiometrically calibrated data may further enhance the accuracy and reliability of forest component segmentation using MS-LiDAR data.

## Acknowledgements

The research is funded by European Spatial Data Research (EuroSDR). Moreover, this project has received funding from the European Union – NextGenerationEU instrument and is funded by the Research Council of Finland under grant number 353264, and UNITE Flagship (337656, 348644, 359404) and Scan4est research infrastructure (346382).